# Dual-wavelength generation and tuning by controlling the apodized grating depth in microring resonators

IS Amiri [1,2*], Volker J. Sorger [3], MM Ariannejad [4], H Ahmad [4], P Yupapin [1,5]

[1] Computational Optics Research Group, Ton Duc Thang University, Ho Chi Minh City, Vietnam;
[2] Faculty of Applied Sciences, Ton Duc Thang University, Ho Chi Minh City, Vietnam
* E-mail: irajsadeghamiri@tdt.edu.vn
[3] Department of Electrical and Computer Engineering, The George Washington University, Washington, D.C. 20052, USA
[4] Photonics Research Centre, University of Malaya, 50603 Kuala Lumpur, Malaysia
[5] Faculty of Electrical & Electroniangcs Engineering, Ton Duc Thang University, District 7, Ho Chi Minh City, Vietnam

**Abstract**— Here we show a photonic design for tunable dual-wavelength generation deploying optical nonlinear mode coupling of two coupled III-V semiconductor microring resonators (MRRs) connected to a pump and drop waveguide buses. Here one of the two rings contains a grating, while the other has a planar surface. The underling mechanism for the dual wavelength generation originates from the resonance-detuning of the spectra resulting in non-linear mode mixing. Tunability of the wavelengths is achieved by altering the grating depth of the MRR and the power coupling coefficients. For the grating design of the MRR we select a trapezoidal-profiled apodized grating to gain low reflectivity at sidelobes. A time-domain travelling wave (TDTW) analysis gives a InGaAsP core refractive index of 3.3 surrounded by a grating InP cladding with n=3.2. We further confirm that the propagation of a Gaussian pulse input with 10 mW power and bandwidth of 0.76 ps is well confined within the mode propagation of the system. Taken together our results show a 2:1 fan-out of two spectrally separate signals for compact and high functional sources on chip.

**Keywords**: Microring resonator (MRR), Apodized grating, InGaAsP/InP semiconductor, dual-wavelength

## 1. Introduction

Multichannel fiber Bragg grating (FBG) filter has been widely applied in the dense wavelength division multiplexing (DWDM) systems, owing to its powerful capacity offering a high number of channels of identical spectral performance for wavelength filtering [1, 2]. Generation of dual-wavelength using passive and active semiconductor waveguides has attracted many research interests recently [3, 4]. They are well known to be useful for applications in the field of optical fiber sensors [5, 6] Zhou et al. have proposed a stable dual-wavelength laser based on cascaded fiber Bragg gratings [7], whereby the laser wavelengths are determined by the fiber Bragg gratings (FBG) and thus the wavelength spacing is fixed by the center wavelengths of the FBGs pair. Another configuration of tunable dual-wavelength is demonstrated in [8], wherein a high birefringence FBG stabilizes a dual-wavelength laser produced when the polarization state of each wavelength is altered using a polarization controller. A large number of researchers have investigated the distinctive properties of photonic crystal fibers, such as wide range single mode operation, dispersion flexibility, large mode area and its application in multi-wavelength generation [9-12]. In most cases, these properties were proven to be wavelength dependent and equivalent to the behavior of wavelength-selective filters [13, 14].

Microring resonator (MRR) based optical mirrors and band-limited reflectors have been the subject of much investigations in recent years, primarily as a result of advancements in fabrication technology and miniaturization of integrated devices [15-17]. MRRs fabricated using planar waveguide technology are well suitable for synthesizing optical filters due to their high-contrast spectral sensitivity, in particular when lossless materials are used [18, 19].

Due to the resulting high quality (Q) factors and compact sizes of these waveguide microresonators, low-loss optical add/drop channel filters can be built [20, 21]. Among the existing biological and chemical sensors, sensors based on integrated optical waveguides have been demonstrated to possess a promising performance. These include planar optical-waveguide sensors [22], directional coupler sensors [23], Mach–Zehnder interferometric sensors [24], grating-coupled waveguide sensors [25, 26], and microresonator sensors [27-31]. In order to minimize the smallest detectable wavelength shift, high Q cavities and a low noise detection system is required. A high Q-factor results in narrow spectral peaks. To keep the waveguide single mode and total internal reflection guiding is required to have very small bend radii [32, 33]. In this research, we propose a coupled MRR system consisting of two MMR resonators made of InGaAsP/InP semiconductor, where the drop bus waveguide has a grating in the core area, while the input MRR does not (Fig. 1). To analyze this system, we use the time-domain travelling wave (TDTW) method, which has previously been deployed to model and simulate both passive and active MRRs [34, 35]. TDTW is based on solving advection equations, and the reader is referred to material from Carroll *et al.* detailed in references [36, 37]. This method is applicable to characterize the spectral responses of the optoelectronics devices as followed here. Our solver is PICWave from Photon Design [38]. In the following we study coupled this design-asymmetric MRRs and show functionality to generate two spectrally different (dual) wavelengths. Next, we discuss the underlying physical principle and show the performance sensitivity of the system.

## 2. Proposed Coupled Microring Resonators (MRRs)

The device consists of two coupled MRRs with different group velocities and effective indices; that is the MRR near the input is a regular photonic waveguide-based ring, while the MRR near the drop output is augmented with a polarization sensitive grating. With our aim to generate not a frequency comb, but two specific and selective (on-demand) spectral output wavelengths, we select two coupled resonators, but detuning one from the other by augmenting it with grating (Fig. 1). Each ring generates a frequency comb based on its free-spectral-range (FSR), where the visibility of the fringes depends on the rings' quality (Q) factor. The grating detunes the respective resonance peaks from the two MRRs, such that they can interfere. Upon constructive interference, the signal mixed output is generated, while deconstructive interference reflects the signal at $\kappa_2$, which effectively reflects the signal. Therefore a periodic resonance can be generated for each reflection spectrum. The grating section is providing the means to suppress repetition of the reflection spectrum and will allow oscillations at a limited number of specific frequencies only [39]. Therefore, a comb of resonances can be generated resulting from the Fourier transform of the created periodic grating in the MRRs structure. Since the optical mode overlaps with the grating, this modal interference is design-tunable; for instance, selecting a different medium as top cladding or altering the different grating periods causes generation of different resonance spacing. Such tunability is interesting for laser sources, and other spectral sensitive applications, as further discussed in the conclusion section below [35, 40].

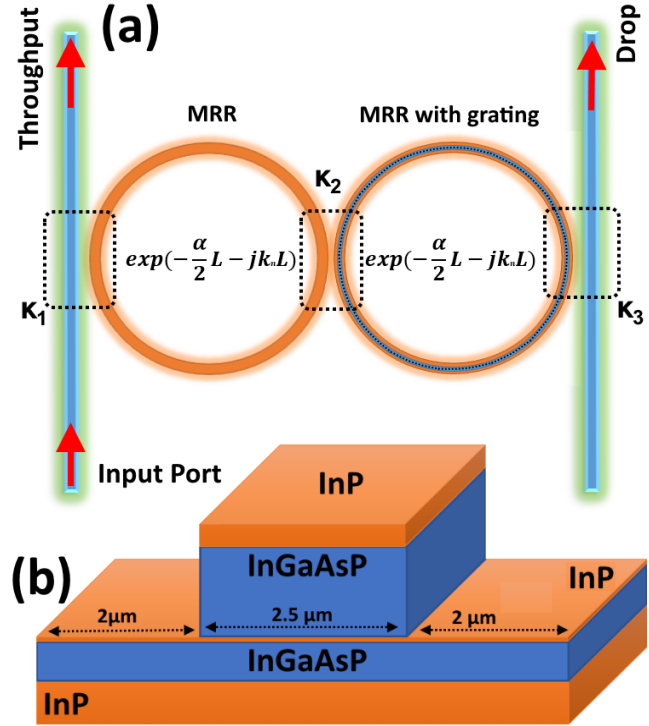

**Fig. 1.** (a) Schematic diagram of the MRR semiconductor structure, each of the MRR has circumference of 50 μm and the MRR coupled to the drop bus waveguide has grating in the core area, (b) InGaAsP/InP profile

The MRRs are made of InGaAsP/InP semiconductor with InGaAsP core having refractive index of 3.31 surrounded by InP (*n*=3.18). The used semiconductor layers sequence (from top to bottom) is shown in Table 1,

**Table 1:** MRR semiconductor layers sequence (from top to bottom)

| Material | Thickness | Refractive index |
|---|---|---|
| InP (cap cladding) | 0.2 μm | 3.18 |
| InGaAsP | 0.84 μm | 3.31 |
| InP (etch stop layer) | 0.02 μm | 3.39 |
| InGaAsP | 0.38 μm | 3.31 |
| InP (substrate) | 0.4 μm | 3.18 |

The input optical field ($E_i$) of the Gaussian pulse is given by [41]

$$E_i(z=0,t) = \sqrt{P_0}\, \exp\!\left[-\frac{t^2}{2T_0^2}\right], \qquad (1)$$

where, the pulse width $T_0$ (related to the pulse full width at half maximum (FWHM) by $T_{\text{FWHM}} \approx 1.665\, T_0$) increases with $z$ (the pulse broadens) according to

$$T(z) = \left[1 + \left(\frac{z}{L_D}\right)^2\right]^{1/2} T_0, \qquad (2)$$

and, consequently, the peak power changes, due to group velocity dispersion (GVD), are given by:

$$P_{(z)} = \frac{P_0}{\left[1 + \left(\frac{z}{L_D}\right)^2\right]^{1/2}} \qquad (3)$$

In Equations 2 and 3, the quantity $L_D = T_0^2/|\beta_2|$ is the dispersion length [42] of the Gaussian pulse and $\beta_2$ is the second order coefficient term of the Taylor expansion of the propagation constant [43, 44]. When Gaussian pulse propagates within the coupled MRRs, the resonant output is formed for each round-trip [45, 46]. Both MRRs have same radius of 8 μm, and the power coupling coefficients are $\kappa_1=\kappa_2=\kappa_3=0.1$. The linear absorption coefficient is $\alpha=0.5(dBmm^{-1})$ and the effective area is 3.03 μm². Widely tunable laser diodes are important components as high-speed light source for wavelength-division-multiplexing (WDM) optical networks and future optical switching application. We have applied an apodized grating on the MRR with total circumference of 50 μm (each MRR's radius is 8 μm). The grating applied to the MRR has a trapezoidal profile as shown in Figure 2. The trapezoid style may be used to define useful grating shapes including square, triangular and sawtooth. Mathematically, the grating period can be calculated according to the Equation (4) [47].

$$\lambda / n_{eff} = 2\Lambda \qquad (4)$$

Where, $n_{eff}$ is the mode effective index, λ is the Bragg wavelength and $\Lambda$ is the grating period. In this research, $n_{eff}$ =3.28 and λ=1.55 μm, therefore the grating period is $\Lambda$=0.236 μm.

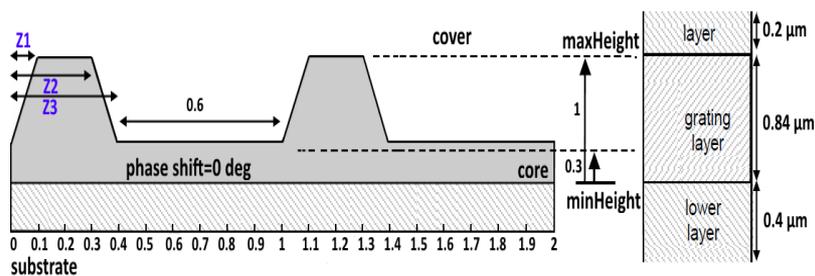

**Fig. 2.** Z position/period, schematic showing the cross section of one period of a grating structure for a trapezoidal style grating shape, where the grating period is 0.236 μm (center wavelength is 1.55 μm), Z1=0.1, Z2=0.3, Z3=0.4 are all fractional values of the grating period, the refractive index below the grating surface is that of the grating layer and the index above the surface is that of the layer above.

The fabrication of apodized Bragg gratings has raised much interest because of its reduced reflectivity at sidelobes. It therefore increases the quality of optical filters and improves the dispersion compensation by simultaneously reducing the group-delay ripples. The side-band reflection peaks can be problematic for many applications, causing crosstalk in WDM systems, instabilities in Q-switched fiber lasers and linewidth broadening in high power fiber lasers. To eliminate these unwanted side-bands it is necessary to fabricate gratings with an apodized profile, where the grating strength varies as a function of length and is typically weaker at both ends of the grating. Apodized gratings offer significant improvement in side-lobe suppression while maintaining reflectivity and a narrow bandwidth. In practice, one apodizes the reflection spectrum by gradually increasing and then decreasing the grating strength along the waveguide. For nonphotosensitive materials the grating is formed by varying the waveguide dimensions, and the grating's depth and duty cycle determine its strength. The fabrication process for apodized gratings must allow one or both of these parameters to vary [48].

The introduced model allows gratings to be apodized, and also allows phase shifts to be introduced between gratings in adjacent sections. In this research, dual-wavelength can be generated with different fraction of the minheight (grating depth), therefore, different grating depth can be applied to the coupled MRRs system. Different grating depth causes the dual-wavelength generated at the throughput and drop ports have different spacing range from 0.2 to 1.43 nm. The changes in dual-wavelength spacing causes this system act as sensor device with sensitivity which is depending on the refractive index of the top cover cladding. In this research, MRRs with a grating section are used to generate comparative outputs using a very compact system and ability to make the proposed system in chip.

## 3. Result and Discussion

The propagation of the input pulse (with 10 mW power, FWHM of 0.76 ps and center wavelength of 1.55 µm) within the MRR semiconductor is shown in Figure 3.

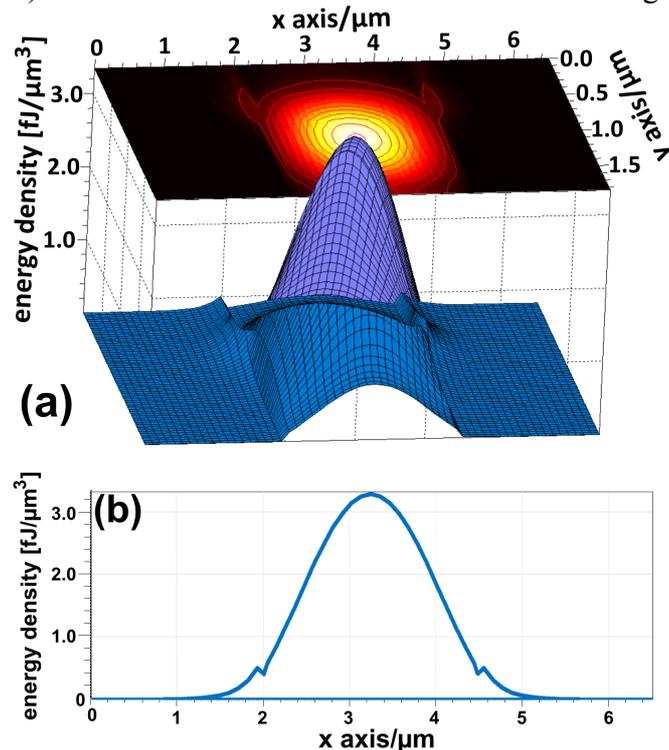

**Fig. 3.** (a) The propagating mode profile of the input pulse within the MRRs semiconductor, where the effective cross section of the mode propagation is 3.03 µm$^2$ and the effective index is 3.28, (b) mode profile of the waveguide with sloped sidewall

As we can see from the Figure 3, there is a good confinement of the input light inside the MRR waveguide. The throughput port outputs of the MRR is shown in Figure 4. The dual-wavelength with tunable spacing within a range of 0.2 nm (26 GHz) to 1.43 nm (178 GHz) could be generated by the variation of the minheight fraction or grating depth on the MRR coupled to the drop bus waveguide.

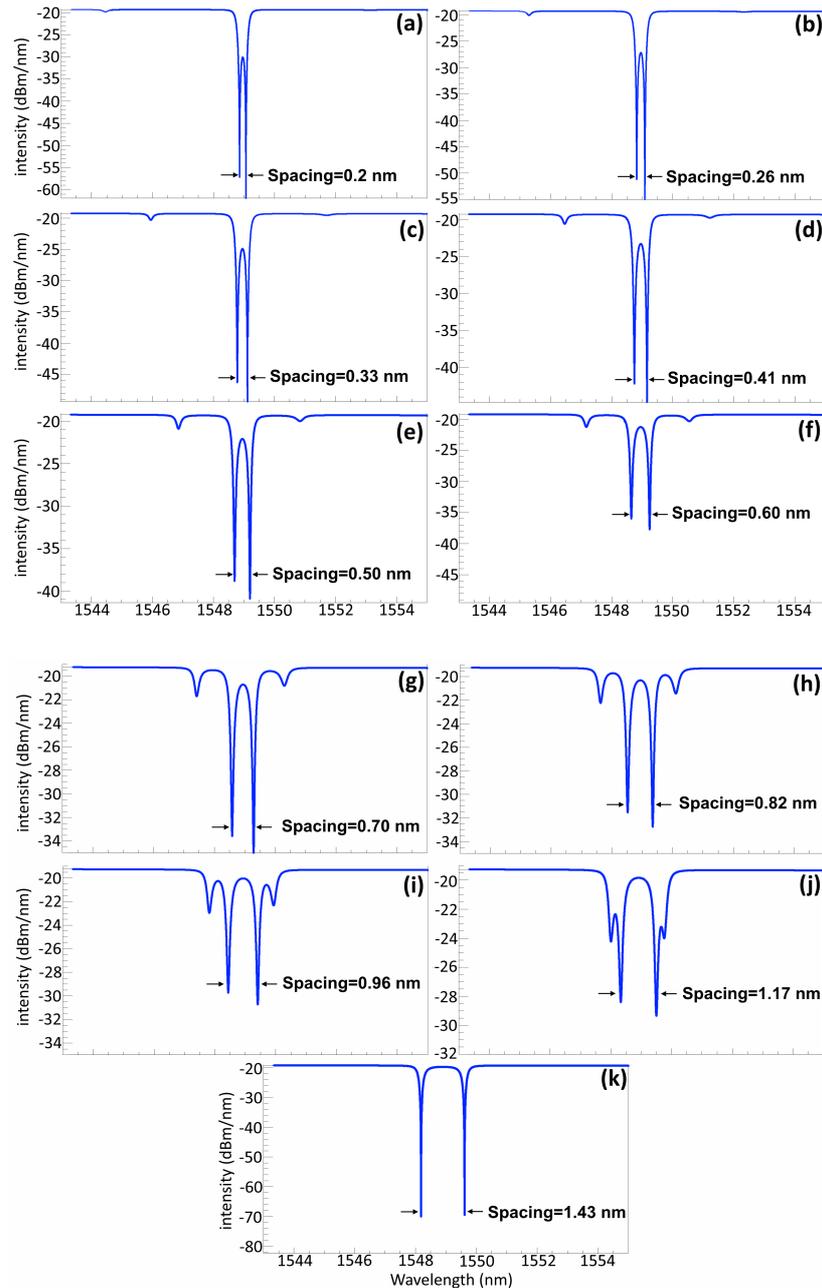

**Fig. 4.** Throughput port outputs, (a) minheight=0, (b) minheight=0.1, (c) minheight=0.2, (d) minheight=0.3, (e) minheight=0.4, (f) minheight=0.5, (g) minheight=0.6, (h) minheight=0.7, (i) minheight=0.8, (j) minheight=0.9, (k) minheight=1 (no grating applied)

The fraction varies between 0 and 1, thus if it is "1" this means that no grating has been applied to the MRR. Figure 5 shows the dual-wavelength spacing versus minheight fraction (grating depth) for the throughput port outputs.

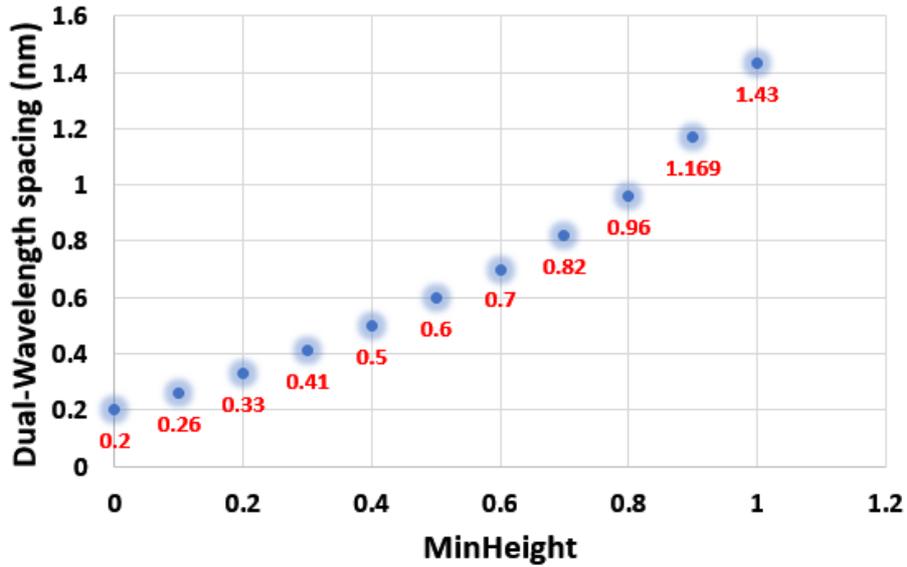

**Fig. 5.** Dual-wavelength spacing versus minheight fraction (grating depth) for the throughput port outputs

Therefore, when there is no grating applied to the MRR, the spacing is maximum and it decreases when grating applied depends on the minheight fraction. Figure 6 shows the dual-wavelength spacing versus power coupling coefficient ($\kappa_1=\kappa_2=\kappa_3$) of the MRRs which varies between 0.1 and 0.9 for the throughput outputs. In this case the minheight fraction is selected to 0.3.

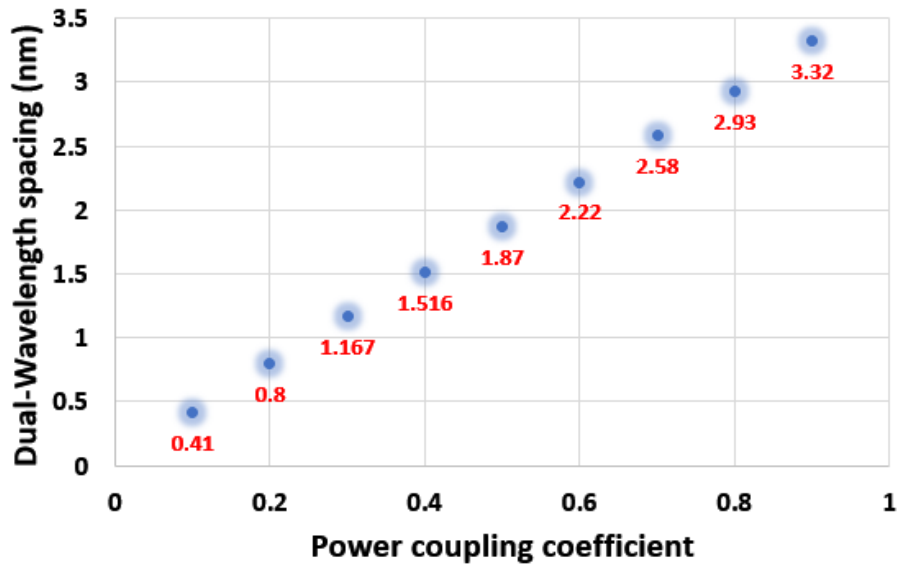

**Fig. 6.** Dual-wavelength spacing versus power coupling coefficient of the MRRs (throughput outputs)

Figure 7 shows the drop port outputs. The multi wavelengths could be generated. The pulse centered at 1549 nm shows a dual-wavelength, owing different dual spacing with different minheight fractions as 0.4, 0.6, 0.8 and 1, therefore the spacing of the dual-wavelength varies between 0.46 and 1.43 nm.

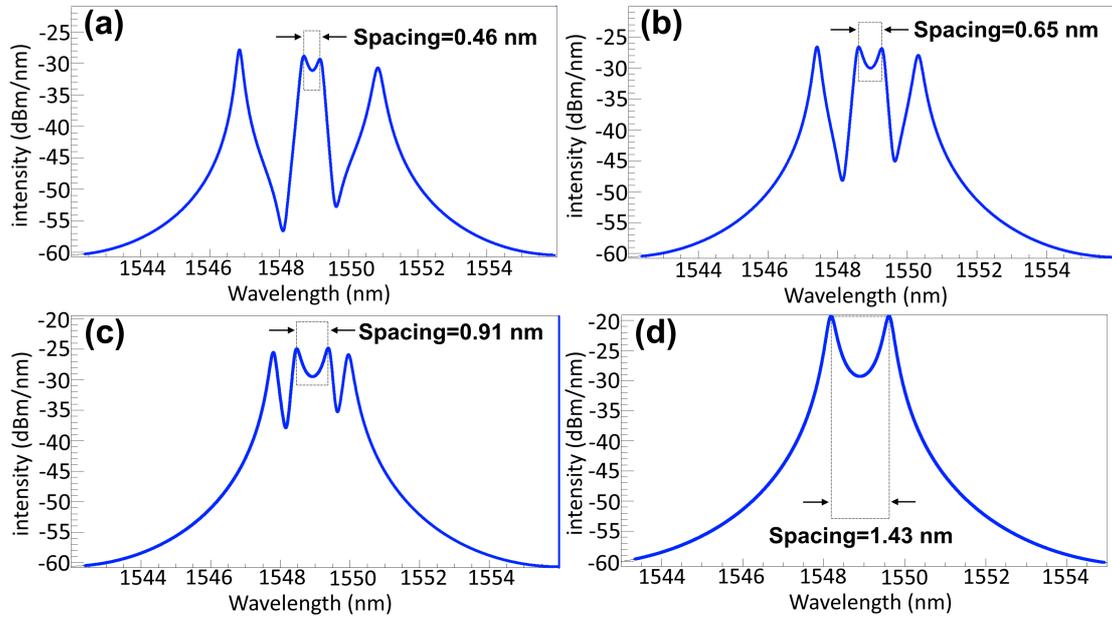

**Fig. 7.** Drop port outputs, (a) minheight=0.4, (b) minheight=0.6, (c) minheight=0.8, (d) minheight=1

The dual-wavelength could be generated at the drop port output signals. When the intensity difference between the two 'wavelengths' in each dual-wavelength is insignificant (less than 1 dB), this is just considered as a single pass band of the filter. Therefore, the minheight fraction has been selected between 0.4 and 1. Figure 8 shows the dual-wavelength spacing versus minheight fraction for the drop port outputs.

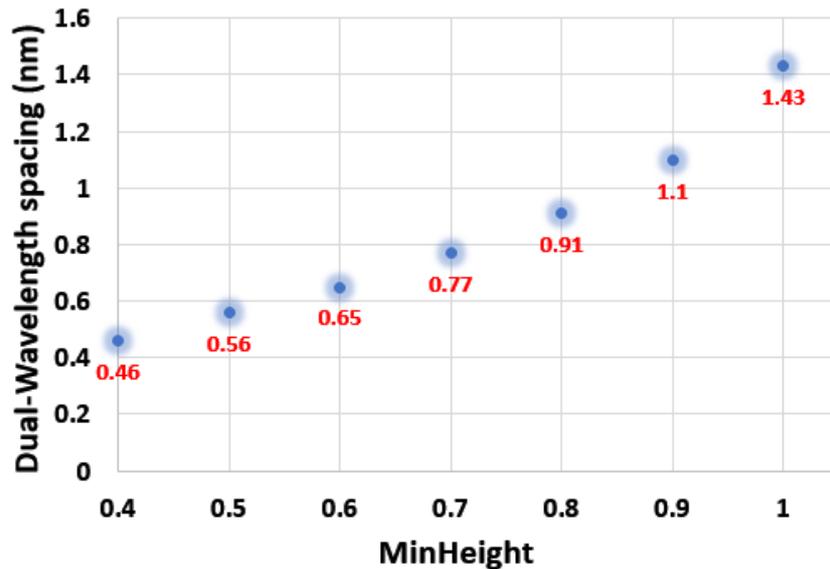

**Fig. 8.** Dual-wavelength spacing versus minheight fraction for the drop port outputs

## 4. Conclusion

This system is able to generate dual-wavelength with tunable spacing in the range of GHz in frequency domain. Further study can be done on this topic based on generating of millimeter wave with GHz center frequencies using the generated dual-wavelengths. By beating the closely center wavelengths of the dual-wavelength, we can obtain center frequencies in millimeter wave range corresponds to that of spacing. The system of InGaAsP/InP coupled semiconductor MRRs owing apodized grating sections has been used to generate a dual-wavelength with tunable spacing ranges from 0.2 nm (corresponds to 26 GHz) to 1.43 nm (corresponds to 178 GHz). The InGaAsP/InP coupled MRRs is consisting of two MRRs, where the MRR coupled to the drop bus waveguide has grating in the core area. The total circumference of each MRR is 50 μm. The simulated results are based on the time-domain travelling wave (TDTW) method. We investigate the propagation of an input Gaussian pulse with a 10 mW power and a bandwidth of 0.76 ps within the proposed coupled MRRs system. This system is able to generate dual-wavelength with spacing in the range of GHz in frequency domain. Stable, single-mode, dual-wavelength lasers have important applications in fiber sensors. The beat between the two laser wavelengths will generate the millimeter wave which can be used in optical communications. In this research the spacing of the generated dual-wavelength varies to the grating depth which is presented by the minheight fraction. The different profile of the grating will affect the outputs of the coupled MRRs, and consequently, will varies the spacing between the two center wavelengths in a dual-wavelength. Spectral tunability of cavities is an enabling functionality for next generation integrated nanophotonic devices and circuits to include light sources and emitters [49-51], reconfigurable switches and modulators [52-59], surface-sensitive devices [60], and application in on-chip data communication [61,62] ideally with nanoscale footprint for high electro-optic device efficiencies [63,64]. For instance, the MRRs either can therefore be used as sensor devices based on an evanescent wave interacting with the surrounding cladding of the system made of semiconductors or be used to generate millimeter wave used in optical communications [61,62].

## Acknowledgements

V.J.S. is supported by the U.S. Air Force Office of Scientific Research-Young Investigator Program under grant FA9550-14-1-0215, and under grant FA9550-14-1-0378 and FA9550-15-1-0447.

## References

[1]     H. Ahmad, N. Saat, and S. Harun, "S-band erbium-doped fiber ring laser using a fiber Bragg grating," *Laser Physics Letters,* vol. 2, p. 369, 2005.
[2]     M. Zulkifli, N. Hassan, N. Awang, Z. Ghani, S. Harun, and H. Ahmad, "Multi-wavelength fiber laser in the S-band region using a Sagnac loop mirror as a comb generator in an SOA gain medium," *Laser Physics Letters,* vol. 7, p. 673, 2010.
[3]     K.-S. Lee and C. Shu, "Stable and widely tunable dual-wavelength continuous-wave operation of a semiconductor laser in a novel Fabry-Perot grating-lens external cavity," *Quantum Electronics, IEEE Journal of,* vol. 33, pp. 1832-1838, 1997.
[4]     U. Keller and A. C. Tropper, "Passively modelocked surface-emitting semiconductor lasers," *Physics Reports,* vol. 429, pp. 67-120, 2006.
[5]     R. A. Potyrailo, S. E. Hobbs, and G. M. Hieftje, "Optical waveguide sensors in analytical chemistry: today's instrumentation, applications and trends for future development," *Fresenius' journal of analytical chemistry,* vol. 362, pp. 349-373, 1998.
[6]     S. Harun, M. Shirazi, and H. Ahmad, "Multiple wavelength Brillouin fiber laser from injection of intense signal light," *Laser Physics Letters,* vol. 4, pp. 678-680, 2007.

[7]  P. Zhou, X. L. Wang, Y. X. Ma, K. Han, and Z. J. Liu, "Stable all-fiber dual-wavelength thulium-doped fiber laser and its coherent beam combination," *Laser Physics,* vol. 21, pp. 184-187, Jan 2011.

[8]  S. Liu, F. P. Yan, W. J. Peng, T. Feng, Z. Dong, and G. K. Chang, "Tunable Dual-Wavelength Thulium-Doped Fiber Laser by Employing a HB-FBG," *Ieee Photonics Technology Letters,* vol. 26, pp. 1809-1812, Sep 15 2014.

[9]  P. S. J. Russell, "Photonic-crystal fibers," *Journal of lightwave technology,* vol. 24, pp. 4729-4749, 2006.

[10]  H. Ahmad, M. R. K. Soltanian, C. H. Pua, M. Alimadad, and S. W. Harun, "Photonic crystal fiber based dual-wavelength Q-switched fiber laser using graphene oxide as a saturable absorber," *Applied Optics,* vol. 53, pp. 3581-3586, 2014/06/01 2014.

[11]  H. Ahmad, M. Soltanian, M. Alimadad, and S. Harun, "Stable narrow spacing dual-wavelength Q-switched graphene oxide embedded in a photonic crystal fiber," *Laser Physics,* vol. 24, p. 105101, 2014.

[12]  M. Soltanian, H. Ahmad, C. Pua, and S. Harun, "Tunable microwave output over a wide RF region generated by an optical dual-wavelength fiber laser," *Laser Physics,* vol. 24, p. 105116, 2014.

[13]  W. Chen, S. Lou, S. Feng, L. Wang, H. Li, T. Guo*, et al.*, "Switchable multi-wavelength fiber ring laser based on a compact in-fiber Mach-Zehnder interferometer with photonic crystal fiber," *Laser physics,* vol. 19, pp. 2115-2119, 2009.

[14]  H. Ahmad, M. R. K. Soltanian, C. H. Pua, M. Z. Zulkifli, and S. W. Harun, "Narrow Spacing Dual-Wavelength Fiber Laser Based on Polarization Dependent Loss Control," *Photonics Journal, IEEE,* vol. 5, pp. 1502706-1502706, 2013.

[15]  B. Liu, A. Shakouri, and J. E. Bowers, "Wide tunable double ring resonator coupled lasers," *Photonics Technology Letters, IEEE,* vol. 14, pp. 600-602, 2002.

[16]  J. K. Poon, J. Scheuer, and A. Yariv, "Wavelength-selective reflector based on a circular array of coupled microring resonators," *IEEE Photonics Technology Letters,* vol. 16, pp. 1331-1333, 2004.

[17]  I. Chremmos and N. Uzunoglu, "Reflective properties of double-ring resonator system coupled to a waveguide," *Photonics Technology Letters, IEEE,* vol. 17, pp. 2110-2112, 2005.

[18]  IS Amiri, SE Alavi, MRK Soltanian, N Fisal, ASM Supa'at, and H Ahmad, "Increment of Access Points in Integrated System of Wavelength Division Multiplexed Passive Optical Network Radio over Fiber," *Scientific Reports,* vol. 5, 2015.

[19]  M Soltanian, IS Amiri, SE Alavi, and H Ahmad, "All Optical Ultra-Wideband Signal Generation and Transmission Using Mode-locked laser Incorporated With Add-drop Microring Resonator (MRR)," *Laser Physics Letters,* vol. 12, 2015.

[20]  R. Grover, V. Van, T. Ibrahim, P. Absil, L. Calhoun, F. Johnson*, et al.*, "Parallel-cascaded semiconductor microring resonators for high-order and wide-FSR filters," *Lightwave Technology, Journal of,* vol. 20, pp. 900-905, 2002.

[21]  V. J. Sorger, Z. Ye, R. F. Oulton, Y. Wang, G. Bartal, X. Yin*, et al.*, "Experimental demonstration of low-loss optical waveguiding at deep sub-wavelength scales," *Nature Communications,* vol. 2, p. 331, 2011.

[22]  T. Okamoto, M. Yamamoto, and I. Yamaguchi, "Optical waveguide absorption sensor using a single coupling prism," *JOSA A,* vol. 17, pp. 1880-1886, 2000.

[23]  B. Luff, R. Wilson, D. Schiffrin, R. Harris, and J. Wilkinson, "Integrated-optical directional coupler biosensor," *Optics letters,* vol. 21, pp. 618-620, 1996.

[24]  Z.-m. Qi, N. Matsuda, K. Itoh, M. Murabayashi, and C. Lavers, "A design for improving the sensitivity of a Mach–Zehnder interferometer to chemical and biological measurands," *Sensors and Actuators B: Chemical,* vol. 81, pp. 254-258, 2002.

[25]  R. Horváth, H. C. Pedersen, N. Skivesen, D. Selmeczi, and N. B. Larsen, "Optical waveguide sensor for on-line monitoring of bacteria," *Optics letters,* vol. 28, pp. 1233-1235, 2003.

[26]  R. Horvath, H. C. Pedersen, N. Skivesen, D. Selmeczi, and N. B. Larsen, "Monitoring of living cell attachment and spreading using reverse symmetry waveguide sensing," *Applied Physics Letters,* vol. 86, p. 071101, 2005.

[27]     E. Krioukov, D. Klunder, A. Driessen, J. Greve, and C. Otto, "Integrated optical microcavities for enhanced evanescent-wave spectroscopy," *Optics letters,* vol. 27, pp. 1504-1506, 2002.

[28]     E. Krioukov, D. Klunder, A. Driessen, J. Greve, and C. Otto, "Sensor based on an integrated optical microcavity," *Optics letters,* vol. 27, pp. 512-514, 2002.

[29]     F. Vollmer, D. Braun, A. Libchaber, M. Khoshsima, I. Teraoka, and S. Arnold, "Protein detection by optical shift of a resonant microcavity," *Applied Physics Letters,* vol. 80, pp. 4057-4059, 2002.

[30]     S. Arnold, M. Khoshsima, I. Teraoka, S. Holler, and F. Vollmer, "Shift of whispering-gallery modes in microspheres by protein adsorption," *Optics letters,* vol. 28, pp. 272-274, 2003.

[31]     C.-Y. Chao and L. J. Guo, "Biochemical sensors based on polymer microrings with sharp asymmetrical resonance," *Applied Physics Letters,* vol. 83, pp. 1527-1529, 2003.

[32]     K. De Vos, I. Bartolozzi, E. Schacht, P. Bienstman, and R. Baets, "Silicon-on-Insulator microring resonator for sensitive and label-free biosensing," *Optics express,* vol. 15, pp. 7610-7615, 2007.

[33]     H. Huang, K. Liu, B. Qi, and V. J. Sorger, "Re-analysis of single-mode conditions for Silicon rib waveguides at 1550 nm wavelength," *Journal of Lightwave Technology,* vol. 34, pp. 3811-3817, 2016.

[34]     L. Wu, J.-J. He, and D. Gallagher, "Modeling of widely tunable V-cavity semiconductor laser using time-domain traveling-wave method," *JOSA B,* vol. 32, pp. 309-317, 2015.

[35]     SE Alavi, IS Amiri, MRK Soltanian, R Penny, ASM Supa'at, and H Ahmad, "Multiwavelength generation using an add-drop microring resonator integrated with InGaAsP/InP sampled grating distributed feedback (SG-DFB)," *Chinese Optics Letters,* vol. 14, p. 021301, 2016.

[36]     L. Zhang, S. Yu, M. Nowell, D. Marcenac, J. Carroll, and R. Plumb, "Dynamic analysis of radiation and side-mode suppression in a second-order DFB laser using time-domain large-signal traveling wave model," *IEEE Journal of Quantum Electronics,* vol. 30, pp. 1389-1395, 1994.

[37]     C. Tsang, D. Marcenac, J. Carroll, and L. Zhang, "Comparison between'power matrix model'and'time domain model'in modelling large signal responses of DFB lasers," *IEE Proceedings-Optoelectronics,* vol. 141, pp. 89-96, 1994.

[38]     "PICWave, Photon Design Inc., http://www.photond.com/products/picwave/picwave_applications_00.htm."

[39]     C. Vázquez and O. Schwelb, "Tunable, narrow-band, grating-assisted microring reflectors," *Optics Communications,* vol. 281, pp. 4910-4916, 2008.

[40]     C. L. Arce, K. De Vos, T. Claes, K. Komorowska, and P. Bienstman, "SOI Microring Resonator Sensor Integrated on a Fiber Facet," in *Lab-on-Fiber Technology*, ed: Springer, 2015, pp. 53-68.

[41]     G. P. Agrawal, *Nonlinear fiber optics*: Academic press, 2007.

[42]     I. S. Amiri, S. E. Alavi, S. M. Idrus, A. S. M. Supa'at, J. Ali, and P. P. Yupapin, "W-Band OFDM Transmission for Radio-over-Fiber link Using Solitonic Millimeter Wave Generated by MRR," *IEEE Journal of Quantum Electronics,* vol. 50, pp. 622 - 628, 2014.

[43]     MRK Soltanian, IS Amiri, WY Chong, SE Alavi, and H Ahmad, "Stable dual-wavelength coherent source with tunable wavelength spacing generated by spectral slicing a mode-locked laser using microring resonator," *IEEE Photonics Journal,* vol. 7, 2015.

[44]     H Ahmad, MRK Soltanian, IS Amiri, SE Alavi, AR Othman, and ASM Supa'at, "Carriers Generated by Mode-locked Laser to Increase Serviceable Channels in Radio over Free Space Optical Systems," *IEEE Photonics Journal,* vol. 7, 2015.

[45]     IS Amiri, SE Alavi, MRK Soltanian, H Ahmad, N Fisal, and ASM Supa'at, "Experimental Measurement of Fiber-Wireless (Fi-Wi) Transmission via Multi Mode Locked Solitons from a Ring Laser EDF Cavity," *IEEE Photonics Journal,* vol. 7, 2015.

[46]     S. E. Alavi, I. S. Amiri, S. M. Idrus, ASM Supa'at, J. Ali, and P. P. Yupapin, "All Optical OFDM Generation for IEEE802.11a Based on Soliton Carriers Using MicroRing Resonators " *IEEE Photonics Journal,* vol. 6, 2014.

[47]     K. O. Hill and G. Meltz, "Fiber Bragg grating technology fundamentals and overview," *Journal of lightwave technology,* vol. 15, pp. 1263-1276, 1997.

[48]  J. Hastings, M. H. Lim, J. Goodberlet, and H. I. Smith, "Optical waveguides with apodized sidewall gratings via spatial-phase-locked electron-beam lithography," *Journal of Vacuum Science & Technology B,* vol. 20, pp. 2753-2757, 2002.

[49]  N. Li, et al., "Nano III-V Plasmonic Light-Sources for Monolithic Integration on Silicon", Nature: Scientific Reports, 5, 14067 (2015).

[50]  K. Liu, V. J. Sorger, "An electrically-driven Carbon nanotube-based plasmonic laser on Silicon" Optical Materials Express, 5, 1910-1919 (2015).

[51]  K. Liu, N. Li, D. K. Sadana, V. J. Sorger, "Integrated nano-cavity plasmon light-sources for on-chip optical interconnects" ACS Photonics**, 3, 233-242** (2016).

[52]  S. K. Pickus, S. Khan, C. Ye, Z. Li, and V. J. Sorger, "Silicon Plasmon Modulators: Breaking Photonic Limits" IEEE Photonic Society 27, 6 (2013).

[53]  K. Liu, Z.R. Li, S. Khan, C. Ye, V. J. Sorger, "Ultra-fast electro-optic modulators for high-density photonic integration" Laser & Photonics Review, 10, 11-15 (2015).

[54]  Z. Ma, Z. Li, K. Liu, C. Ye, V. J. Sorger, "Indium-Tin-Oxide for High-performance Electro-optic Modulation", Nanophotonics, 4, 1 (2015).

[55]  R. Amin, C. Suer, Z. Ma, J. Khurgin, R. Agarwal,

[56]  V. J. Sorger, "Active Material, Optical Mode and Cavity Impact on electro-optic Modulation Performance" arXiv:1612.02494 (2016).

[57]  R. Amin, C. Suer, Z. Ma, J. Khurgin, R. Agarwal, V. J. Sorger, "A Deterministic Guide for Material and Mode Dependence of On-Chip Electro-Optic Modulator Performance", Solid-State Electronics, Special Issue, DOI: 10.1016/j.sse.2017.06.024 (2017).

[58]  Z. Ma, M. H. Tahersima, S. Khan, V. J. Sorger, "Two-Dimensional Material-Based Mode Confinement Engineering in Electro-Optic Modulators," IEEE Journal of Selected Topics in Quantum Electronics, vol. 23, no. 1, 1-8 (2017).

[59]  J. K. George, V. J. Sorger, "Graphene-based Solitons for Spatial Division Multiplexed Switching", Optics Letters 42, 4, 787-790 (2017).
[60] M. H. Tahersima, et al. "Testbeds for Transition Metal Dichalcogenide Photonics: Efficacy of Light Emission Enhancement in Monomer vs. Dimer Nanoscale Antennas", ACS Photonics, 4, 1713-1721 (2017).

[60]  M. H. Tahersima, at al., "Testbeds for Transition Metal Dichalcogenide Photonics: Efficacy of Light Emission Enhancement in Monomer vs. Dimer Nanoscale Antennas", ACS Photonics, 4, 1713-1721 (2017).

[61]  S. Sun, A. Badaway, T. El-Ghazawi, V. J. Sorger, "Photonic-Plasmonic Hybrid Interconnects: Efficient Links with Low latency, Energy and Footprint", IEEE Photonics, 7, 6, 15 (2015).

[62]  V. K. Narayana, S. Sun, A.-H. Badawya, V. J. Sorger, T. El-Ghazawi, "MorphoNoC: MorphoNoC: Exploring the Design Space of a Configurable Hybrid NoC using Nanophotonics," Microprocessors and Microsystems (2017).

[63]  A. Fratalocchi, et al., "Nano-optics gets practical: Plasmon Modulators", Nature Nanotechnology, 10, 11-15 (2015).

[64]  K. Liu, A. Majumdar, V. J. Sorger, "Fundamental Scaling Laws in Nanophotonics" Scientific Reports, 6, 37419 (2016).